\documentclass[twocolumn,preprintnumbers,amsmath,amssymb,superscriptaddress,pre]{revtex4}

\usepackage{subfigure}
\usepackage{wrapfig}
\usepackage{amsmath,amsfonts,amsthm,amssymb}
\usepackage{enumitem}
\usepackage{graphicx,epstopdf}
\usepackage{color}

\newcommand{\reffig}[1]{Fig.~\ref{#1}}
\newcommand{\refeq}[1]{Eq.~\ref{#1}}

\newcommand{\refsec}[1]{Sec.~\ref{#1}}
\newcommand{\expect}[1]{< {#1} >} 
\newcommand{\Ffive}{F_{5\%}}

\begin{document}

\title{Attention Competition with Advertisement}
\author{Uzay Cetin}
\affiliation{Department of Computer Engineering, Bogazici University, Istanbul}
\affiliation{Department of Computer Engineering, Istanbul Gelisim University, Istanbul}

\author{Haluk O. Bingol}
\affiliation{Department of Computer Engineering, Bogazici University, Istanbul}
\date{\today}

\begin{abstract}
In the new digital age, information is available in large quantities.
Since information consumes primarily the attention of its recipients,
the scarcity of attention is becoming the main limiting factor. 
In this study, we investigate the impact of
advertisement pressure on a cultural market 
where consumers have a limited attention capacity.
A model of competition for attention is developed and 
investigated analytically and by simulation. 
Advertisement is found to be much more effective 
when attention capacity of agents is extremely scarce. 
We have observed that 
the market share of the advertised item improves
if dummy items are introduced to the market
while the strength of the advertisement is kept constant.
\end{abstract}

\maketitle

\section{Markets}	
Traditionally every product or service has
a price tag.
In order to get it, 
one has to pay the price.
Nowadays, 
the price of items in some markets becomes so low, 
even to the point of free-of-charge, 
that this concept of ``pay-to-get'' is challenged,
especially in the era of Internet.
It is quite a common fact that 
one can get many products and  services
paying absolutely nothing.
Among these are
internet search (Google, Yahoo), 
email (Gmail, Hotmail),
storage (DropBox, Google, Yahoo), 
social networks (Facebook, Twitter, LinkedIn),
movie storage (Youtube),
communication (Skype, WhatsApp), 
document formats (PDF, RTF, HTML),
various software platforms (Linux, LaTeX, eclipse, Java)
and
recent trend in education (open course materials and massive open online courses (MOOC)). 

Companies providing services, 
where their users pay no money at all, 
is difficult to explain in Economics.
Even if these products are free to its user,
there is still a sound business plan behind them.
To obtain a large market share is the key in their business plan 
as in the cases of 
Google, Facebook, LinkedIn, or Skype.
Once they become widely used,
the company 
starts to use its customer base to create money.

\subsection{New market concepts}

In order to understand such markets new concepts  
such as two-sided markets and attention economy are developed.
In a \emph{two-sided market}, a company acts as a bridge between 
two different type of consumers~\cite{
	Parker2005MS}.
It provides two products: 
one is free and 
the other with a price.
Free products are used to capture the attention.
Products with price are used to monetize this attention.
A set of very interesting examples of two-sided markets
	including 
	credit cards, 
	operating systems,
	computer games,
	stock exchanges, 
can be found in ref~\cite{
	Parker2005MS}.

Suppose there are many competing products at the free side of a two-sided market.
In theory, a customer can get all the products available.
In practice, this is hardly the case.
Abundance
of immediately available products 
can easily exceed customers capacity to 
consume them.
One way to look at this phenomenon is that
products compete for the attention of the users, 
which is referred as \emph{attention economy} 
in the literature~\cite{
	Davenport2002IBJ,
	Weng2012SF,
	Wu2007PNAS}.

Attention scarcity 
due to the
vast amount of immediately available products
is also the case for cultural markets. 
In a \emph{cultural market},
it is assumed to have an infinite supply for cultural products
and 
it is assumed that individual consumption behaviour is not independent of other's consumption decisions
\cite{Salganik10022006, 
	Herdagdelen2008IJMPC}.

\subsection{Compulsive markets}
We focus on markets, 
that are slightly different,
where customer compulsively purchases the item once he is aware of it.
Clearly, 
this kind of compulsive buying behavior cannot happen 
for high priced items such as cars or houses.
On the other hand, 
it could be the case 
for relatively low priced items such as movie DVDs or music CDs. 
This pattern of ``compulsive purchasing'' behavior becomes clearly acceptable,
if the items become free 
as in the case of 
web sites, 
video clips,  
music files, and
free softwares, 
especially free mobile applications.
There are a number of services that provide such items including 
Youtube, Sourceforge, AppStore.

We will call such markets as \emph{compulsive markets} and
we consider the dynamics of the consumers rather then the economics of it. 
These new kind of markets call for new models.
In this work, 
the Simple Recommendation Model of 
ref~\cite{
	Bingol2008PRE, 
	Bingol2005}
is extended to such a model.
We use the extended model to answer the following questions:
Under which conditions advertisement mechanism outperforms the recommendation process? 
How much advertisement is enough to obtain certain market share?
We first present our analytic approach and then compare it with 
simulation results.

\section{Background}

A compulsive buyer becomes aware of a product in two ways:
(i)~By local interactions within his social network, 
i.e. by means of word-of-mouth.
(ii)~By global interactions,
i.e. by means of advertisement.

Word-of-mouth recommendations by friends make products socially contagious. 
Research on social contagion can provide answers to the question of how things become popular. 
Gladwell states, 
"Ideas, products, messages and behaviors spread like 
viruses do"~\cite{
	Gladwell2000Book}. 
He claims that 
the best way to understand the emergence of fashion trends is 
to think of them as epidemics. 
Infectious disease modeling is also useful 
for understanding opinion formation dynamics. 
Specifically, the transmission of ideas within a population is treated as if it were the transmission of an infectious disease. 
Various models have been proposed to examine this 
relationship~\cite{
	Bass1969MS,
	Rogers2003Book,
	Goldenberg2001ML,
	Dodds2005JTB,
	Herdagdelen2008IJMPC,
	PastorSatorras2001PRL,
	Bingol2008PRE}. 
There exist recent works 
whose essential  assumption is the fact that 
an old idea is never repeated once 
abandoned~\cite{
	Bornholdt2011PRL,
	Kondratiuk2012PRE}. 
In other words, 
agents become immune to older ideas 
like in the susceptible-infected-recovered (SIR) model. 
However, behaviors, trends, etc, can occur many times over and over again. 
In this case it can be modeled as 
susceptible-infected-susceptible (SIS) model.
In completely different context, 
limited attention and its relation to income distribution is 
investigated~\cite{
	Banerjee2008}.

\subsection{Epidemic spreading}
\label{sec:EpidemicSpreading}

The study of how ideas spread is often referred to as 
social contagion~\cite{
	EasleyKleinberg2010Book}. 
Opinions can spread from one person to another like diseases. 
An agent is called \emph{infected} iff 
it has the virus.
It is called \emph{susceptible} iff 
it does not have the virus.

Using the SIS model of epidemics,
the system can be modeled as a Markov chain. 
Consider a population of $N$ agents.
Let $S_i$ be the state in which the number of infected agents is $i$.
The state space is composed of $ N+1 $ states, 
$\{S_{0}, S_{1}, \ldots,S_{N} \}$ 
with $S_{0}$ and  $S_{N}$ being the reflecting boundaries.
The system starts with the state $S_{0}$ where nobody is infected.

Let $\mathbf{T} = [t_{ij}]$ be 
the $(N+1) \times (N+1)$ transition matrix of the Markov chain 
where 
$t_{ij}$ is the transition probability from state $S_{i}$ to state $S_{j}$.
As a result of a single recommendation,
there are three possible state transitions:
The number of infected agents can 
increase or decrease by one or 
stay unchanged.
Such a system is called birth death process~\cite{
	Ross2009Book}. 
Hence, 
$\mathbf{T}$ is a
tridiagonal matrix with entries given as
\begin{align*}
	t_{ij} = 
	\begin{cases}
		p_{i}, &j = i + 1, \\
		l_{i}, &j = i, \\
		q_{i}, &j = i - 1, \\
		0, &\text{otherwise}
	\end{cases}
\end{align*}
where $p_{i}, l_{i}$ and $q_{i}$ are the transition probabilities.

Then the stationary distribution 
$\boldsymbol{\pi} = [\pi_{0} \cdots \pi_{N}]^{\top}$ 
of the Markov chain can be obtained from its transition 
matrix~\cite{
	Ross2009Book} 
which satisfies 
\begin{equation}
	\pi_{i} 
	= \prod_{k=1}^i \frac{p_{k-1}}{q_k}\pi_{0}
	\quad
	\text{ and }
	\quad
	\sum_{i=0}^N \pi_{i}=1.
\label{eq:stationaryDistribution}
\end{equation}

\subsection{Simple Recommendation Model}

The \emph{Simple Recommendation Model} (\emph{SRM}) 
reveals the relation 
between 
the fame and
the memory size of the 
agents~\cite{
	Bingol2008PRE, 
	Bingol2005}.
The SRM investigates 
how individuals become popular 
among agents with limited memory size 
and analyzes the word-of-mouth effect in its simplest form. 
The SRM differs from many previous models by its emphasis on the scarcity of memory. 
In the SRM, 
agents, 
that have a strictly constant memory size $M$, 
learn each other solely via recommendations.

A \emph{giver} agent selects an agent, that he knows, and 
\emph{recommends} to a \emph{taker} agent.
Since memory space is restricted to $M$,
the taker \emph{forgets} an agent to make space for the \emph{recommended} one.
This dynamics is called a \emph{recommendation} which 
is given more formally in \refsec{sec:model}. 
Note that 
(i)~The selections have no sophisticated mechanisms.
All selections are made uniformly at random.
(ii)~Any agent can recommend to any other agent. 
Therefore underlining network of interactions is a complete graph.
(iii)~Taker has to accept the recommended,
that is,
he has no options to reject.

In the SRM, 
no agent initially is different than the other.
So the initial fames of agents are set to be the same
where \emph{fame} of an agent is defined as 
the ratio of the population that knows the agent.
Recommendations break the symmetry of equal fames.
As recommendations proceed, 
a few agents get very high fames
while the majority of the agents get extremely low fames, 
even to the level of no fame at all.
Once an agent's fame becomes $0$,
that is, 
\emph{completely forgotten},
there is no way for it to come back.
In the limit, 
the system reaches an \emph{absorbing state} 
where exactly $M$ agents are known by every one, i.e. fame of $1$,
and the rest becomes completely forgotten, i.e. fame of $0$.
The SRM offers many possibilities for extension.
It is applied to minority communities living in a 
majority~\cite{
	Delipinar2009}.
A recent work extends forgetting mechanism by introducing 
familiarity~\cite{
	Yi-Ling2013}.

\section{Proposed Model}

In SRM, 
(i)~the spread of information through out the system 
is managed by recommendation only
and 
(ii)~the results are obtained by 
simulations~\cite{
	Bingol2005,
	Bingol2008PRE}. 
In this article, 
we propose \emph{Simple Recommendation Model with Advertisement} 
(\emph{SRMwA}) 
that extends SRM in the following ways:
(i)~In addition to recommendation,  
advertisement pressure as new dynamic is introduced. 
(ii)~Moreover, an analytical approach is developed as well as simulations. 
Distinctively, by SRMwA, we investigate the conditions under which 
social manipulation by advertisement overcomes pure recommendation.

\subsection{New interpretation for SRM}

In the original model of SRM, 
agents recommend other agents
and
the term of memory size is used for the number of agents one can 
remember~\cite{
	Bingol2005,
	Bingol2008PRE}. 
As one agent is known more and more by other agents, his fame increases.
In the extended model of SRMwA, agents recommend items rather than agents.
Since items consume the limited attention of agents,
there is a competition among items for attention. 
For these reasons,
we prefer to use the term of ``attention capacity'' in spite of the term memory size
for the number of information an agent can handle.
The focus of the work is no longer the fame of the agents 
but the attention competition among items.

Note that the proposed model allows us to consider items in a wider sense.
Rather than a unique object such as 
Mona Lisa of Leonardo, 
we consider items that are easily reproduced so that 
there are enough of them for everybody to have, 
if they wanted to.
Therefore items are not only products and services 
but also as political ideas, fashion trends, or cultural products 
as in the case of ref~\cite{
	Herdagdelen2008IJMPC}.

\subsection{Advertisement}

We extend the SRM to answer the following question:
What happens if some items are deliberately promoted?  
Suppose a new item, denoted by $a$, is
\emph{advertised} to the over-all population.
At each recommendation, 
the taker has to select between 
the recommended item $r$
and the advertised one $a$.
The item that is selected by the taker is 
called the \emph{purchased item}, denoted by $\beta$.

\subsection{Model}
\label{sec:model}

Adapting the terminology of SRM~\cite{
	Bingol2008PRE} 
to SRMwA, 
a \emph{giver} agent $g$ recommends an item, 
that she already owns, to an individual.
The item and the individual are called 
the \emph{recommended} $r$ and the \emph{taker} $t$, respectively. 
The taker pays attention to, 
that is, \emph{purchases}, 
either the recommended or the advertised item.
When the attention capacity becomes exhausted, 
in order to get space for the purchased item,
an item $f$ that is already owned by the taker is \emph{discarded}.
The \emph{market share} of an item 
is defined to be the ratio of population that owns the item.

The SRMwA is formally defined as follows.
Let 
$\mathcal{N} = \{ 1,2,\ldots,N\}$ and 
$\mathcal{I} = \{ 1,2,\ldots,I\}$ be 
the sets of agents and items, respectively. 
Let 
$g, t \in \mathcal{N}$ and 
$r, f, \beta \in \mathcal{I} \cup \{a\}$  
represent 
the \emph{giver} and 
the \emph{taker} 
agents, 
the \emph{recommended},
the \emph{discarded} and 
the \emph{purchased} 
items, respectively.

The attention ``stock'' of an agent $i$, denoted by $m(i)$,  
is the set of distinct items 
that $i$ owns.
We say agent $i \in \mathcal{N}$ \emph{owns} item $j \in \mathcal{I}$ 
iff $j \in m(i)$. 
For the sake of simplicity, 
we assume 
that all agents have the same \emph{attention capacity} $M$,
that is, 
$|m(i)| = M$ for all $i \in \mathcal{N}$.
The attention capacity of an agent is limited in the sense that 
no one can pay attention to the entire set of items
but to a small fraction of it,
that is, $M \ll I$. 
Instead of directly using $M$, 
we relate $M$ to $I$ by means of
\emph{attention capacity ratio},
defined as 
$\rho = M / I$.
Since $0 \le M \le I$,
we have $0 \le \rho \le 1$.

The recommendation and advertisement dynamics compete.   
The taker agent select either the recommended or the advertised item as the purchased one.
Let the \emph{advertisement pressure}, $p$,
be the probability of selecting the advertised item as the purchased item.

The modified recommendation is composed of the following steps: 

\begin{enumerate}[itemsep=-4pt, topsep=4pt, partopsep=0pt]

	\item[i)] 
	$g$ is selected.

	\item[ii)] 
	$t$ is selected.

	\item[iii)] 
	$r \in m(g)$ is selected by $g$ for recommendation.

	\item[iv)] 
	$t$ selects $\beta$ where 
	$\beta$ is set 
	to $a$  with probability $p$, and
	to $r$ with probability $1-p$.

	\item[v)] 
	The recommendation stops if $\beta$ is already owned by $t$.
		
	\item[vi)] 
	Otherwise, 
	$f \in m(t)$ is selected by $t$ for discarding and
	$\beta$ is put to the space emptied by $f$.
\end{enumerate}
Note that all selections are uniformly at random.
With these changes,
the SRMwA becomes a model for compulsive markets with advertisement.

\subsection{Some special cases}

In general, one expects that 
the market share of the advertised item increases
as advertisement get stronger.
Depending the strength of advertisement,
there are a number of special cases, 
the dynamics of which can be explained without any further investigation.

\begin{enumerate}[itemsep=-4pt, topsep=4pt, partopsep=0pt]

	\item[i)] 
	\textbf{No advertisement.}
	Note that in the case of no advertisement, 
	the original SRM is obtained since
	the purchased item is always the recommended item, 
	i.e. $\beta = r$.
	In this case,
	the advertised item has no chance and its the market share is $0$.

	\item[ii)] 
	\textbf{Pure advertisement.}
	When the taker has no choice but get the advertised one, 
	i.e. $\beta = a$,
	recommendation has no effect.
	In this cases after every agent becomes a taker once,
	the market share of the advertised is $1$.
	Note that in this case 
	the system will stop evolving any further.
	Interestingly, 
	this is a different state than the absorbing states of the SRM.

	\item[iii)] 
	\textbf{Strong advertisement.}
	In the case of very strong advertisement,
	the taker almost always select the advertised item.
	Once all agents have the advertised item,
	the market share of the advertised item is $1$ and
	the system becomes the SRM but with attention capacity of $M-1$.
	
\end{enumerate}

\section{Analytical Approach}
\label{sec:Ana}

Note that SRMwA resembles epidemic spreading.
We explore epidemic spreading to explain SRMwA as far as we can.
Consider the advertised item as a virus. 
Agent $j$ is called \emph{infected} iff 
it has the advertised item in its attention stock, 
that is,
$a \in m(j)$
otherwise
It is called \emph{susceptible} that is $a \notin m(j)$.
Then 
the stationary distribution $\boldsymbol{\pi}$ provides 
the probability of the number of agents owning the advertised item when 
the system operates infinitely long durations.
Hence,
the mean value of the stationary distribution $\boldsymbol{\pi}$ 
reveals our prediction for the number of infected agents.
In other words, 
the expected number of agents that adopted the advertised item is 
the mean value of this distribution.
That is, 
using \refeq{eq:stationaryDistribution}, one obtains
\[
	\expect{\boldsymbol{\pi}}
	= \sum_{i = 0}^{N} i \pi_{i}
	= \pi_{0} \sum_{i=0}^{N} i \prod_{k=1}^{i} \frac{p_{k-1}}{q_{k}}.
\]
Hence, the expected market share of the advertised item becomes
\[
	\expect{F_{a}} 
	= \frac
		{\expect{\boldsymbol{\pi}}} 
		{N} 
\]
where 
$F_{a}$ is the market share of the advertised item. 

\subsection{Calculation of transition probabilities}

\begin{figure}
	\includegraphics[height=\columnwidth,angle=90]{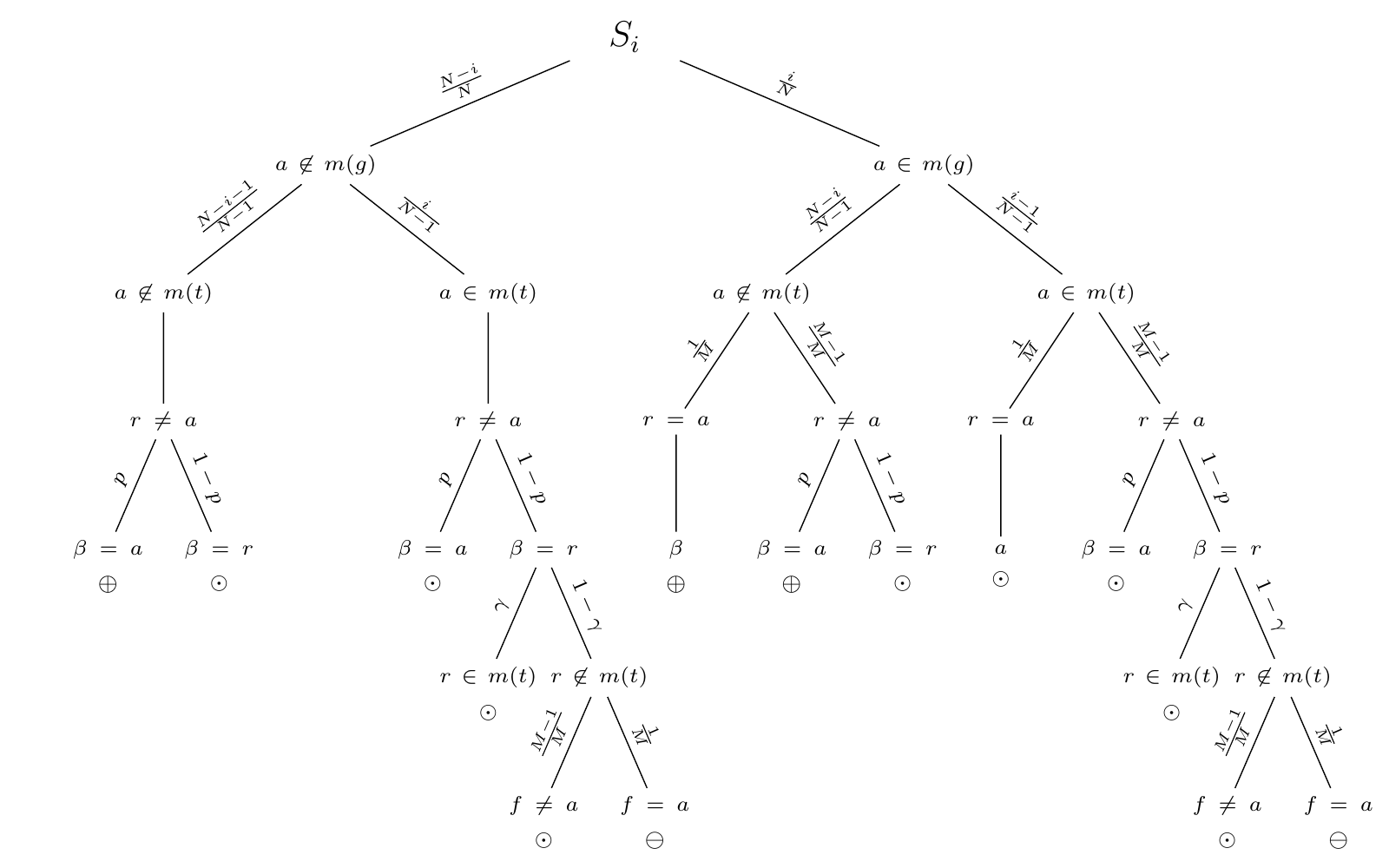}
	\caption{
		Tree diagram for possible selections.
	}
	\label{fig:tree} 
\end{figure}
In order to obtain the expected market share of the advertised item,
we need to figure out the stationary distribution $\boldsymbol{\pi}$, 
which, in turn, calls for transition probabilities $p_{i}, l_{i}$ and $q_{i}$.

Suppose the system is in $S_{i}$
and follow the steps of recommendation process given in \refsec{sec:model}.
The possible selections can be represented by a tree given in \reffig{fig:tree}.
A path starting from the root $S_{i}$ to a leaf in the tree corresponds to a recommendation.
The paths that increase the number of infected agents are
marked by a $\oplus$ sign at the leaf. 
Similarly, recommendations resulting a transition of $S_{i} \rightarrow S_{i-1}$ are marked by a $\ominus$.
The remaining paths that correspond to no state change are marked by a $\odot$.

Note that there three $\oplus$ and two $\ominus$ paths.
Note also that the correspondence between 
the levels in the tree and
the steps of recommendation given in \refsec{sec:model}.
At each level one particular selection is made 
and the corresponding probability is assigned.

\begin{enumerate}[itemsep=-4pt, topsep=4pt, partopsep=0pt]

	\item[i)] 
	$a \in m(g)$ level.
	The first level branching in \reffig{fig:tree} corresponds to 
	the selection of infected or susceptible giver.
	There are $N$ possible agents to be selected as $g$.
	If system is in state $S_{i}$, 
	then the probability of selecting an infected giver is 
	$\frac{i}{N}$.

	\item[ii)] 
	$a \in m(t)$ level.
	The second level branching is due to 
	the selection of infected or susceptible taker.
	Once $g$ is selected,
	there are $N-1$ candidates left for $t$. 
	The probability of selecting an infected taker depends on 
	whether the selected giver is infected or not.
	For example, in the right most path, $g$ is infected.
	So,
	the probability of selecting an infected taker for this case is as
	$\frac{i-1}{N-1}$.  
	 
	\item[iii)]
	$r = a$ level.
	Now consider what the giver recommends. 
	Depending on the path, the giver could be infected and
	could recommend the advertised item.
	Then 
	the probability of an infected giver recommending $a$ is
	$\frac{1}{M}$,
	since there are $M$ items in its stock.
	 
	\item[iv)]
	$\beta = a$ level.
	The fourth level illustrates the taker's purchase decision.
	The taker agent either follows 
	the advertisement 
	with probability $p$
	or he accepts the recommended item 
	with probability $1-p$.
	 
	\item[v)]
	$r \in m(t)$ level.
	Let $\gamma$ be the probability of $r$
	being already owned by the taker agent.
	In this case, 
	the taker agent does not do any changes in her stock.
		 
	\item[vi)]
	$f = a$ level.
	It is possible that $a$ can be chosen to be the forgotten.
\end{enumerate}

The transition probabilities 
can be obtained from \reffig{fig:tree} as
\begin{eqnarray}
	p_{i}  
	&= 
	 &\frac{N-i}{N(N-1)} 
	 \left[
		\left(
			N-1 - \frac{i}{M}
	    \right)
	    p
		+ \frac{i}{M}
	 \right],
	\label{eq:pi}\\
	q_{i}  
	&=
	 &\frac{i (1-p) (1-\gamma)}{N (N-1) M} 
	 \left[
		N-i + \frac{(i-1)(M-1)}{M}
	 \right],
	 \label{eq:qi}\\
	l_{i} 
	&= 
	&1 - (p_{i} + q_{i}).
\end{eqnarray}
Note that 
(i)~These equations satisfy the expected boundary conditions
$q_{0} = 0$, and
$p_{N} = 0$.
(ii)~$p_{i} >0$ for all $i = 0, \cdots, N-1$.
%
(iii)~$q_{i} = 0$ for all $i$ when $p = 1$ or $\gamma = 1$.
Therefore, 
for $p = 1$ or $\gamma = 1$,
the system drifts to $S_{N}$ and stays there forever.


%

\subsection{Discussion on the value of $\gamma$}

The stationary distribution can be obtained by means of
\refeq{eq:stationaryDistribution}, 
\refeq{eq:pi} and 
\refeq{eq:qi}.
The only unknown in these equations is $\gamma$,
which is introduced in the fifth step of recommendation given in \refsec{sec:model}.
$\gamma$ is defined as the probability of recommended item 
to be already owned by the taker agent.
Unfortunately, $\gamma$ cannot be obtained analytically
except for the extreme case of $M=1$.
Therefore, we should find ways to approximate its value.

A first order estimate for $\gamma$ could be $\rho = M / I$,
since taker owns $M$ item out of $I$ in total.
$\gamma$ is close to 1,
when $M$ is in the range of $I$,
since every agent owns almost all the items.
The situation is quite different for $M \ll I$.
Since every item initially has the same market share,
$\gamma$ starts with a small value at the beginning.
As recommendations proceeds,
we know that some items becomes completely forgotten~\cite{
	Bingol2008PRE}.
Therefore $\gamma$ increases as the number of recommendations increase and
becomes $1$ when the systems reaches one of its absorbing state.
In this respect,
$\gamma$ can be interpreted as the degree of closeness to an absorbing state. 
In order to investigate near absorbing state behavior, 
we set $\gamma =  \max \{0.5, M/I \}$ in our analytic results given in \reffig{fig:FigSimRhoAll}~(b) where $0.5$ is arbitrarily selected.

\subsection{Extremely scarce attention capacity}
\label{sec:M1}

For the extremely scarce attention capacity of $M = 1$,
$\gamma$ can be evaluated.
Consider the paths in \reffig{fig:tree}.
For $M = 1$, 
the paths which contain a $(M -1)/M$ edge become 
paths with zero probabilities.
The only non-zero probability path,
involving $\gamma$,
is the one terminating at the left $\ominus$ leaf.
In this path the  giver does not know the advertised item, $a \not \in m(g)$,
while the taker does, $a \in m(t)$.
Since attention capacity is limited to $1$, 
the giver and the taker do own different items.
Therefore, 
the recommended item by the giver cannot be owned by the taker.
Hence,
$\gamma = 0$.

For $M=1$ and $\gamma = 0$, 
the equations \refeq{eq:pi} and \refeq{eq:qi} 
lead to 
\[
	\frac{p_i}{q_{i}}
	= 1 + \frac{N-1}{i} \frac{p}{1-p}
\]
for 
$0 \leq i < N.$
For $p \ne 0$, $p_{i}/q_{i} > 1$.
That means 
for even very small positive advertisement,
the system inevitably drifts to the state $S_{N}$ and
once $S_{N}$ is reached, 
the system stays there forever
since $q_{N} = 0$.
Note that $S_{N}$, 
which corresponds to the state where all agents own the advertised item,
is
the unique absorbing state for this particular case.

\section{Simulation Approach}

In order to simulate the model,
a number of decision have to be made.
The simulations start in such configurations that 
all $I$ items have the same market share 
and
no agent knows the advertised item. 
So that system is initially symmetric with respect to non-advertised items.
When to terminate the simulation is a critical issue.
We set the average number of interactions $\nu = 10^{3}$.
Since there are $N^{2}$ pairwise interactions among agents in both directions, 
the total number of recommendations is set to be $\nu N^{2}$.

(i)~We run our simulation for a population size of $N=100$ 
and an item size of $I=100$.

(ii)~The behavior of the system strongly depends on the attention capacity ratio $\rho$.
We take $\rho$ as a model parameter and run simulation for various values of $\rho$.

(iii)~The advertisement pressure $p$ is another model parameter.
We use $10^{-1}, 10^{-2}, 10^{-3}$ and $10^{-4}$ for $p$.
%

\begin{figure}
\begin{center}
		\includegraphics[width=0.95\columnwidth]{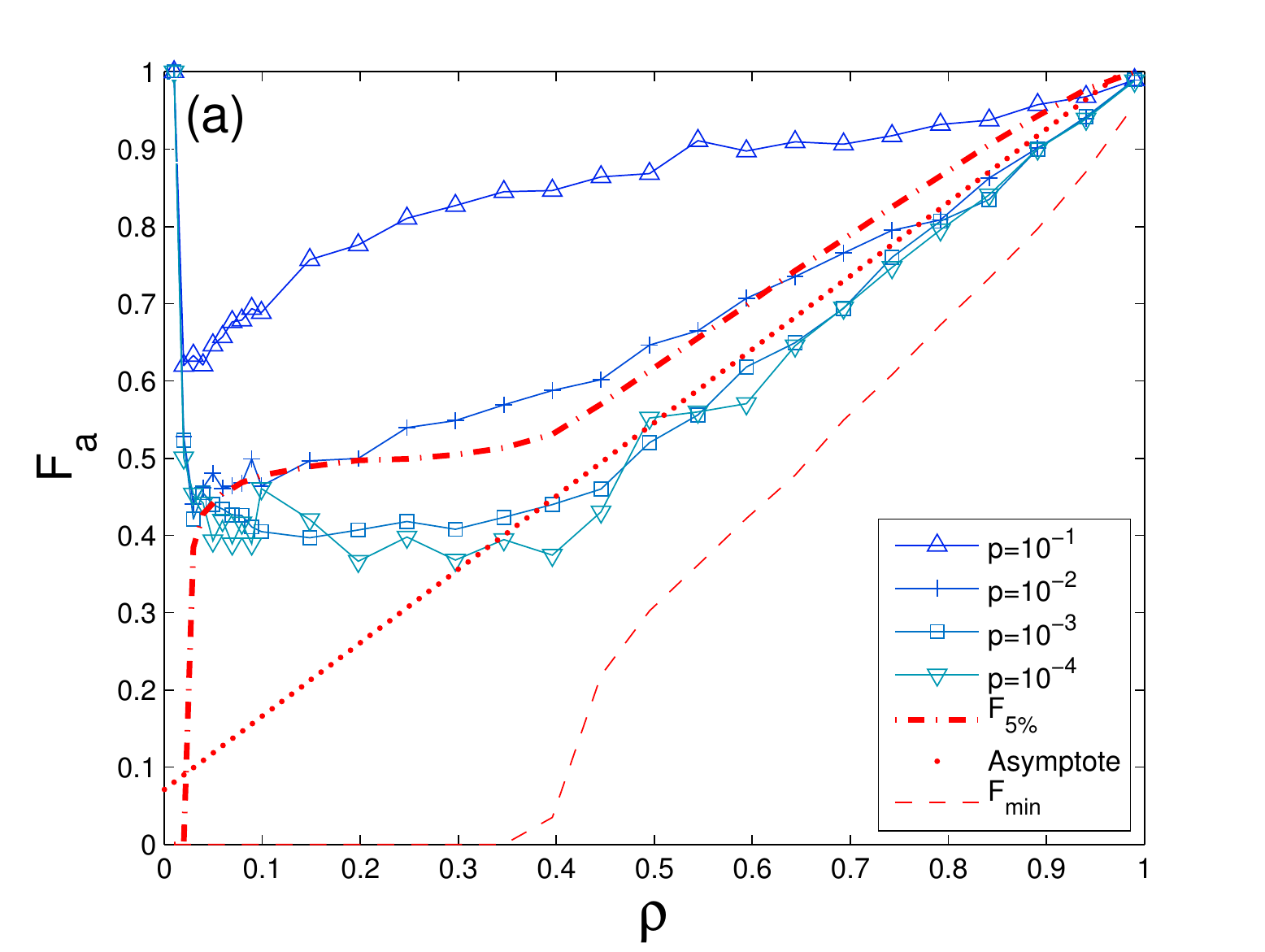}
		\includegraphics[width=0.95\columnwidth]{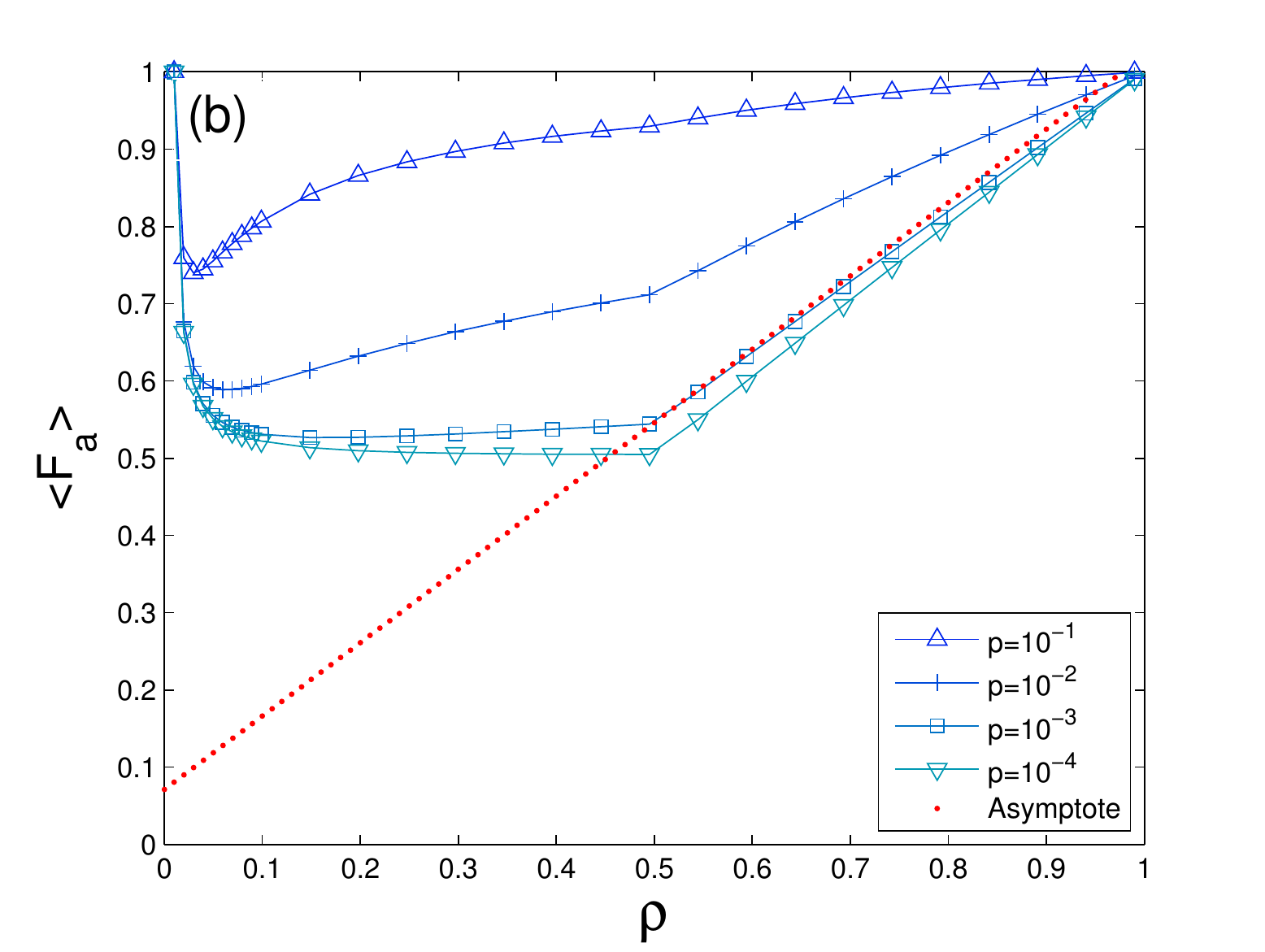}
	\caption{(Color online)
		The market share of advertised item as a function of attention capacity ratio 
		by (a) simulation and (b) analytic approaches.
		$\Ffive$, 
		$F_{min}$ 
		and
		the asymptote line of ref \cite{
			Bingol2008PRE}
		are given for comparison.
	}
	\label{fig:FigSimRhoAll}
\end{center}
\end{figure}

\section{Observations and Discussion}

We investigate the effect of 
the advertisement pressure $p$ and 
the capacity ratio $\rho$ to 
market share $F_{a}$ of the advertised item.
In order to make a quantitative comparison of the simulation results,  
being in the top $5$ percent is arbitrarily set as our criteria. 
Let $F_{5\%}$ 	
denote the lowest market share for an item to be in the top $5$ percent.
Then, 
the advertised item is in the top 5 percent
whenever $F_{a} > \Ffive$.
Let $F_{min}$ be the minimum market share among all the items.

In \reffig{fig:FigSimRhoAll}, 
the simulation results of $F_{a}$,
averaged over 20 realizations and
versus the analytical results of $\expect{F_{a}}$ 
can be seen
for each value of $p \in \{10^{-1}, 10^{-2}, 10^{-3}, 10^{-4} \}$
as functions of $\rho$.
A number of observations can be made:

(i)~The analytic results given in \reffig{fig:FigSimRhoAll}~(b) 
are in agreement with 
the simulation results in \reffig{fig:FigSimRhoAll}~(a). 
Model predictions on $\expect{F_{a}}$ 
can quantitatively reproduce the simulation results of $F_a$
although we use an approximated value for $\gamma$.
We observe that for larger $\nu$, the similarity between analytical and simulation results 
gets even better.

(ii)~The curves of $\Ffive$ in \reffig{fig:FigSimRhoAll}~(a) resemble 
that of in ref~\cite{
	Bingol2008PRE}, 
although advertisement is not the case for the latter.
Line $y = 0.95 x + 0.071$,
which is given as an asymptote 
for $\Ffive$ for large values of $N$ in ref~\cite{
	Bingol2008PRE},
is also plotted in \reffig{fig:FigSimRhoAll}~(a) for comparison purposes.

(iii)~Note that for $\rho < 0.05$,
all $F_{a}$ curves approaches to 1 and 
$\Ffive$ becomes 0.
This is due to finite size effect.
At an absorbing state,
there would be exactly the same $M$ items purchased by all the agents
and the remaining items are completely forgotten.
For $I = 100$, $\rho < 0.05$ means that $M < 5$.
That is, there is no space left for the fifth item.
Hence, 
in near absorbing state, 
the market share of the fifth item, $\Ffive$, approaches to $0$.
On the other hand,
any promotion, 
i.e.  $p > 0$, 
is enough to push the advertised item into 
the top $M$ items.

(iv)~The minimum market share $F_{min}$ becomes $0$,
when at least one item is completely forgotten. 
This occurs for $\rho < 0.35$ in \reffig{fig:FigSimRhoAll}~(a)
which is consistent with ref~\cite{
	Bingol2008PRE}. 
We also observe that for larger $\nu$, 
the advertised item 
leaves smaller share of attention to others, 
that forces the zero crossing of $F_{min}$ to occur
at an higher level of $\rho$.

(v)~As expected,
a strong advertisement, i.e. $p = 10^{-1}$,
easily gets the advertised item into the top $5$ percent 
since $F_{a}$ curve for $p = 10^{-1}$ is always higher than that of $F_{5\%}$ in 
\reffig{fig:FigSimRhoAll}~(a)
while a weak promotion such as $p = 10^{-3}$ or $10^{-4}$ cannot.
The case of $p = 10^{-2} \approx \frac{1}{I + 1}$ for $I = 100$ is interesting.
For small and moderate values of $\rho$, 
i.e. $\rho < 0.6$, 
the advertised item is in the top 5 percent
except for one point.
For the large 
values 
of $\rho$,
this is not the case.

(vi)~How agents allocate their attention,
when the attention capacity becomes a limiting factor?
This is the critical question
for markets of attention economy. 
Consider the extreme case of attention capacity $M = 1$,
which corresponds to $\rho=0.01$ in \reffig{fig:FigSimRhoAll} .
In this case,
surprisingly, even a very small positive value of $p$ is enough 
for the entire population to get the advertised item, 
i.e. $F_{a} = 1$, 
when $M = 1$.
This observation is analytically investigated in \refsec{sec:M1}.

\subsection{Item size effect}

\begin{figure}[b]
	\includegraphics[width=0.95\columnwidth]{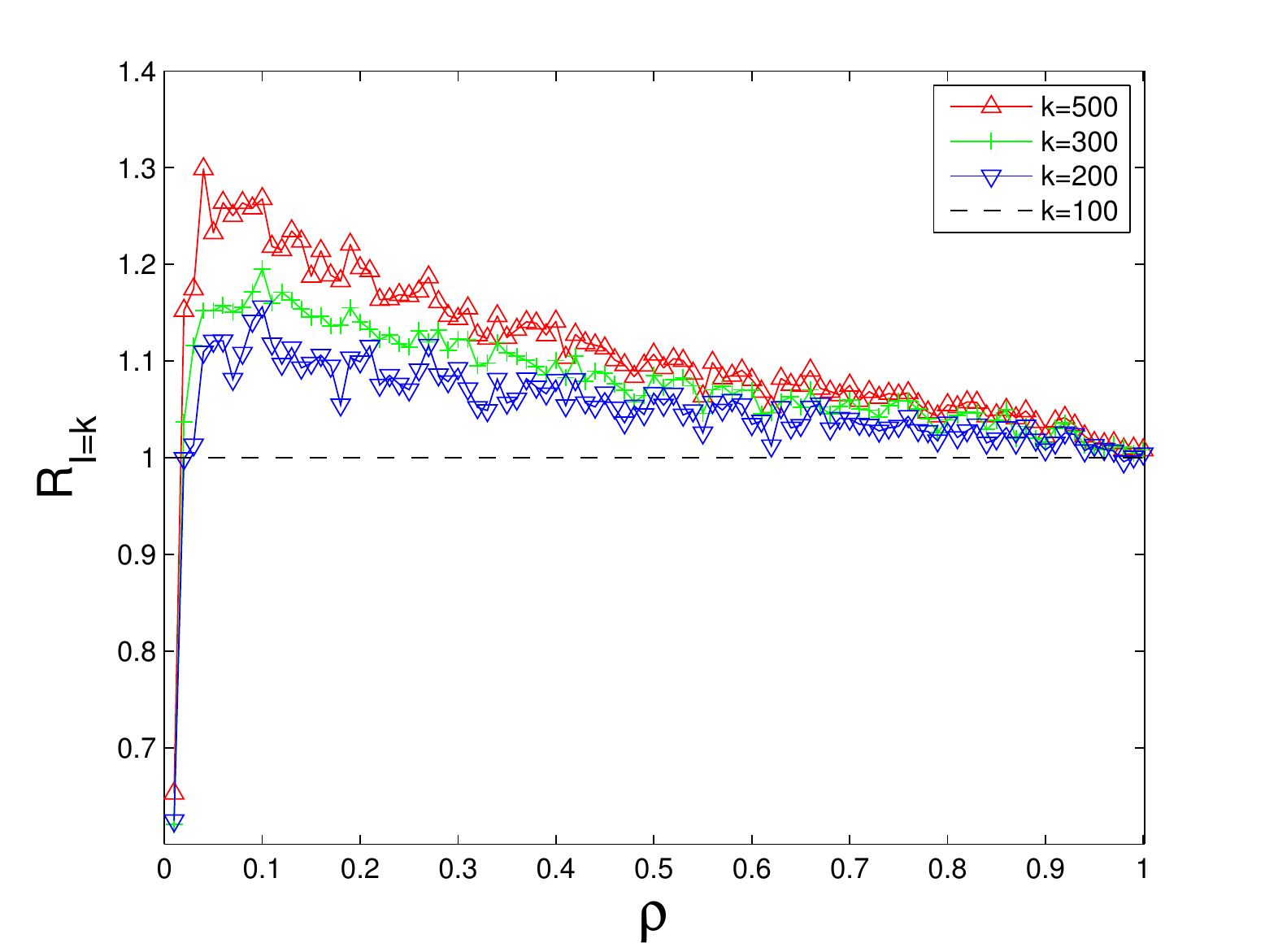}
	\caption{(Color online)
		Effect of item size to the market share of the advertised item
		for $p=10^{-1}$
		is invesitgated as a function of attention capacity ratio.
	}
	\label{fig:ItemEffect}
\end{figure}

We run new simulations with different item sizes of $I$
when $N$ is fixed to $100$.
Let $F_{a}(I = k)$ denote the market share of the advertised item when $I = k$.
Then 
we accept $F_{a}(I = 100)$ as the reference market share and
define relative market share $ R_{I = k}$ with respect to $I = 100$ as follows
\[
	R_{I = k}= \frac{F_{a}(I = k)}{F_{a}(I = 100)}.
\]
In \reffig{fig:ItemEffect}, 
we observe that for all $k \in \{100,200,300,500\}$, 
$R_{I = k} \geq 1$ when $p$ is fixed to $10^{-1}$ except for $\rho=0.01$.
The case of $\rho=0.01$ corresponds to $M=1$ for $I=100$. 
As explained in \refsec{sec:M1}, $F_a$ gets its maximum value of 1, for $M=1$.
That is why, $R_{I = k} \leq 1$ for $\rho=0.01$.


We have observed that the market share of the advertised item improves
while the number of items are increased 
even if the advertisement pressure is kept constant.
In order to push market share up,
increasing the advertisement pressure, is not usully an option in practical life.
This can be an interesting interpretation.
If one cannot increase the intensity of advertisement, i.e $p$, 
it is better to have higher number of items, i.e. $I$.
When that happens, 
the advertised item have better chances to get into the top 5 percent. 
In order to obtain this operating point, 
one may purposefully introduce some dummy items.
This unexpected prediction of the model needs to be further investigated.

\subsection{Closeness to the absorbing state}
\begin{figure}[h]
	\includegraphics[width=0.95\columnwidth]{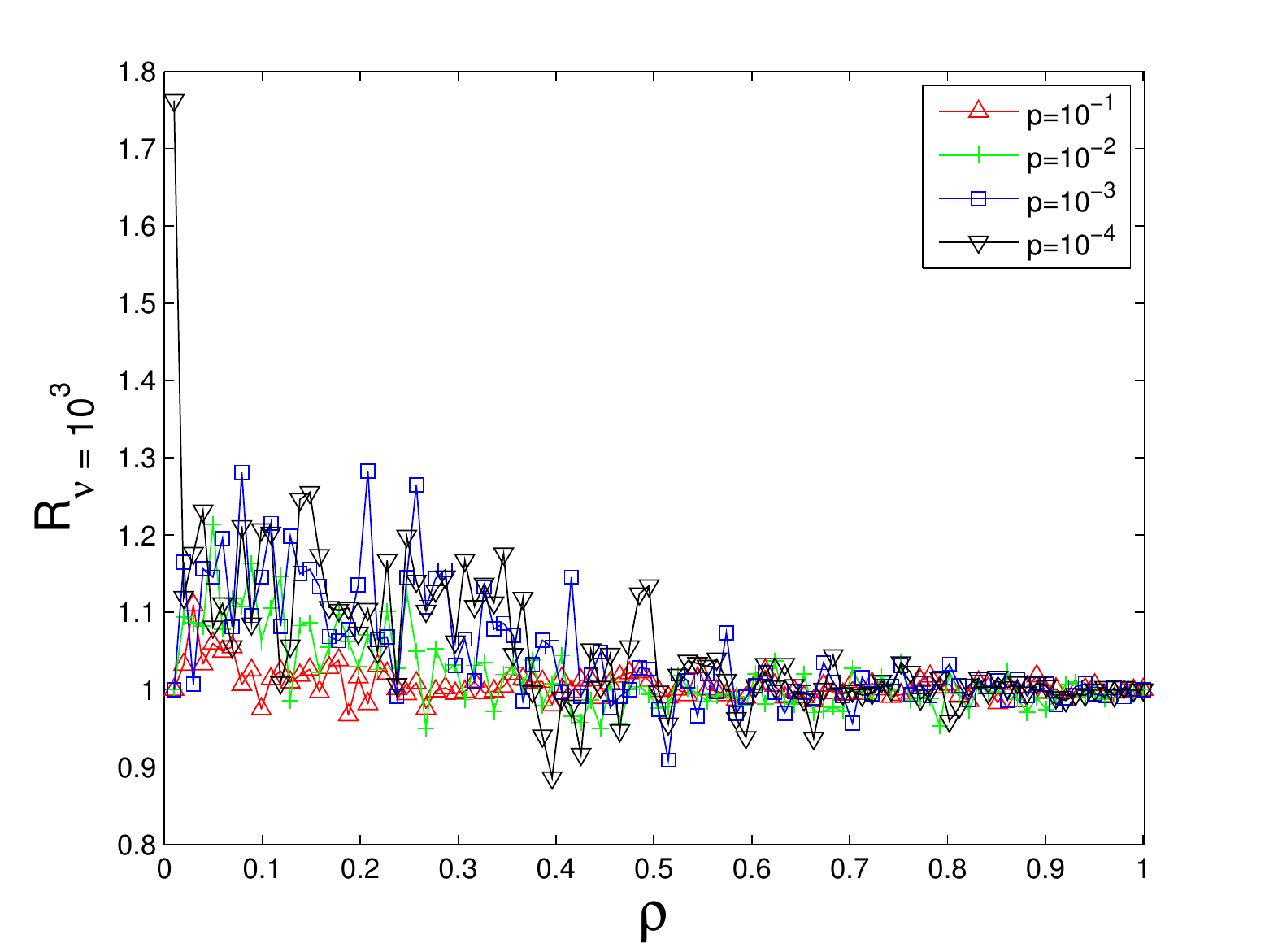}
	\caption{(Color online)
		The relative market share of $R_{\nu = k}$ 
		at $\nu = 10^3$ is invesitgated as a function of attention capacity ratio .
	}
	\label{fig:vEffect}
\end{figure}

The system gets closer to one of its absorbing states as 
the number of recommendations increases 
which is controlled by simulation parameter $\nu$.
Let $F_{a}(\nu = k)$ be the market share of the advertised item 
after $\nu N^{2}$ recommendations.
We define relative market share $R_{\nu = k}$ at $\nu = k$ with respect to $\nu = 10^{2}$ as
\[
	R_{\nu = k} = \frac{F_{a}(\nu = k)}{F_{a}(\nu = 10^{2})}.
\] 
The relative market share at $\nu = 10^{3}$ 
is given 
in \reffig{fig:vEffect}
for different values of $p \in \{10^{-1},10^{-2},10^{-3},10^{-4}\}$
when $N = I = 100$.
%

We consider the system stationary if $R_{\nu = k}$ becomes $1$, 
that is,
the system stops changing with $\nu$.
We observe in \reffig{fig:vEffect} that
as the attention capacity or the advertisement pressure gets higher, 
model becomes closer to the stationarity.
More advertisement pressure is not so different than increasing the number of iterations. 
Both are favorable for the market share of the advertised item.

\section{Conclusions}

The SRM as a model for pure word-of-mouth marketing is studied in 
ref~\cite{
	Bingol2005,
	Bingol2008PRE}. 
We extend the SRM to attention markets with advertisement.
This model constructs a theoretical framework for not only items 
but 
studying the propagation of any phenomena such as ideas or trends
under limited attention. 

The model is investigated analytically and by simulation.
The analytical results agree with the simulations.
As expected, 
strong advertisement forces every one to get the advertised item in all conditions.

Interestingly, 
when the 
attention capacity is small compared to the number of items,
even a very weak advertisement can do the job.
This behavior is analytically shown for the case of $M = 1$ and 
observed in the results of both simulations and analytic calculations 
as $\rho$ approaches to $0$.
This can be interpreted as 
when individuals have limited attention capacity,
they tend to adopt 
what is promoted globally rather than 
recommended locally.
We have also found that introducing more standard items to the market,
is good for the market share of the advertised item.
This observation may lead to interesting political consequences in terms of 
public attention and political administration.
For example, 
public opinion can be kept under control 
by means of increasing the number of issues, 
possibly by means of artificial ones,
so that the  
promoted idea is easily accepted by large audiences.
This prediction calls for further investigation.

In this current work, 
there is a unique advertised item.
The model can be extended to cover more than one promoted items.
All selections are uniformly at random.
One may investigate the effects of some other selection mechanism 
as in the case of ref~\cite{Yi-Ling2013}.
We have a complete graph as the graph of interactions.
One can investigate 
other graphs of interactions such as Scale-Free, Small-World, regular or random graphs.
The structure of interactions can also be improved by introducing a radius of influence.
One may extend the model by introducing
the concept of quality 
for items
or letting agents prefer
some items intrinsically
as in ref
\cite{Herdagdelen2008IJMPC}.

\acknowledgments 

Authors would like to thank to Gulsun Akin for pointing at the two-sided markets.
This work was partially supported 
by Bogazici University Research Fund (BAP-2008-08A105), 
by the Turkish State Planning Organization (DPT) TAM Project (2007K120610),
by TUBITAK (108E218)
and
by COST action MP0801.

\bibliographystyle{ieeetr}
\bibliography{SRMwA.bib}{}

\begin{thebibliography}{10}

\bibitem{Parker2005MS}
G.~G. Parker and M.~W. {Van Alstyne}, ``{Two-Sided Network Effects: A Theory of
  Information Product Design},'' {\em Management Science}, vol.~51, no.~10,
  pp.~1494--1504, 2005.

\bibitem{Davenport2002IBJ}
T.~H. Davenport and J.~C. Beck, ``{The strategy and structure of firms in the
  attention economy},'' {\em Ivey Business Journal}, vol.~66, no.~4,
  pp.~49--54, 2002.

\bibitem{Weng2012SF}
L.~Weng, A.~Flammini, A.~Vespignani, and F.~Menczer, ``{Competition among memes
  in a world with limited attention.},'' {\em Scientific Reports}, vol.~2,
  p.~335, 2012.

\bibitem{Wu2007PNAS}
F.~Wu and B.~A. Huberman, ``{Novelty and collective attention.},'' {\em
  Proceedings of the National Academy of Sciences of the United States of
  America}, vol.~104, no.~45, pp.~17599--17601, 2007.

\bibitem{Salganik10022006}
M.~J. Salganik, P.~S. Dodds, and D.~J. Watts, ``Experimental study of
  inequality and unpredictability in an artificial cultural market,'' {\em
  Science}, vol.~311, no.~5762, pp.~854--856, 2006.

\bibitem{Herdagdelen2008IJMPC}
A.~Herdagdelen and H.~Bingol, ``{A Cultural Market Model},'' {\em International
  Journal of Modern Physics C}, vol.~19, no.~02, p.~271, 2008.

\bibitem{Bingol2008PRE}
H.~Bingol, ``{Fame emerges as a result of small memory},'' {\em Physical Review
  E}, vol.~77, no.~3, p.~036118, 2008.

\bibitem{Bingol2005}
H.~Bingol, ``{Fame as an Effect of the Memory Size},'' {\em Lecture Notes in
  Computer Science}, vol.~3733, pp.~294--303, 2005.

\bibitem{Gladwell2000Book}
M.~Gladwell, {\em The Tipping Point: How little things can make a big
  difference}.
\newblock Little Brown and Company, Boston, 2000.

\bibitem{Bass1969MS}
F.~M. Bass, ``{A New Product Growth for Model Consumer Durables},'' {\em
  Management Science}, vol.~15, no.~5, pp.~215--227, 1969.

\bibitem{Rogers2003Book}
E.~M. Rogers, {\em {Diffusion of Innovations}}.
\newblock Free Press, 5~ed., 2003.

\bibitem{Goldenberg2001ML}
J.~Goldenberg, B.~Libai, and E.~Muller, ``{Talk of the network: A complex
  systems look at the underlying process of word-of-mouth},'' {\em Marketing
  Letters}, vol.~12, no.~3, pp.~211--223, 2001.

\bibitem{Dodds2005JTB}
P.~S. Dodds and D.~J. Watts, ``{A generalized model of social and biological
  contagion.},'' {\em Journal of Theoretical Biology}, vol.~232, no.~4,
  pp.~587--604, 2005.

\bibitem{PastorSatorras2001PRL}
R.~Pastor-Satorras and A.~Vespignani, ``{Epidemic Spreading in Scale-Free
  Networks},'' {\em Physical Review Letters}, vol.~86, no.~14, pp.~3200--3203,
  2001.

\bibitem{Bornholdt2011PRL}
S.~Bornholdt, M.~Jensen, and K.~Sneppen, ``{Emergence and Decline of Scientific
  Paradigms},'' {\em Physical Review Letters}, vol.~106, no.~5, p.~058701,
  2011.

\bibitem{Kondratiuk2012PRE}
P.~Kondratiuk, G.~Siudem, and J.~Ho{\l}yst, ``{Analytical approach to the model
  of scientific revolutions},'' {\em Physical Review E}, vol.~85, no.~6,
  p.~066126, 2012.

\bibitem{Banerjee2008}
A.~Banerjee and S.~Mullainathan, ``{Limited attention and income
  distribution},'' {\em The American Economic Review}, vol.~98, no.~2, 2008.

\bibitem{EasleyKleinberg2010Book}
D.~Easley and J.~Kleinberg, {\em {Networks, Crowds, and Markets: Reasoning
  about a highly connected world}}.
\newblock Cambridge University Press, 2010.

\bibitem{Ross2009Book}
S.~Ross, {\em {Introduction to Probability Models}}.
\newblock Academic Press, 10~ed., 2009.

\bibitem{Delipinar2009}
S.~Delipinar and H.~Bingol, ``{Application of SRM to Diverse Populations},''
  {\em Complex Sciences}, vol.~4, pp.~1063--1071, 2009.

\bibitem{Yi-Ling2013}
W.~Yi-Ling and Z.~Gui-Qing, ``{Optimal convergence in fame game with
  familiarity},'' {\em Chaos, Solitons \& Fractals}, vol.~56, pp.~222--226,
  2013.

\end{thebibliography}

\end{document}